\documentclass{amsart}
\usepackage{amsmath,amscd,amssymb,amsfonts,amsthm,graphicx}

\usepackage{caption}
\usepackage{subcaption}

\theoremstyle{plain}

\theoremstyle{definition}

\numberwithin{figure}{section}

\def\scrm{\scriptsize\rm}

\def\cal{\mathcal}

\def\cS{{\cal S}}

\def\Bbb{\mathbb}

\def\bC{\Bbb C}

\def\bR{\Bbb R}

\def\bZ{\Bbb Z}

\def\ot{\otimes}
\def\ott{{\otimes^2}}

\def\Span{\mathrm{Span}}

\def\thup{{\mbox{\scrm th}}}
\def\tr{\mathrm{tr}}

\begin{document}

\title[Simplicial molecules]{The vibrational modes of simplicial molecules}

\author{Charles H.\ Conley}
\address{Charles H.\ Conley
\\Department of Mathematics
\\University of North Texas 
\\Denton, Texas, 76209, USA} 
\email{conley@unt.edu}

\author{Jon Erickson}
\address{Jon Erickson
\\Department of Mathematics
\\University of California, Davis
\\Davis, California, 95616, USA}
\email{jferickson@ucdavis.edu}

% \keywords{}
% \subjclass{}

\begin{abstract}
Consider a ``simplicial molecule'': $n$ equal point masses placed at the vertices of a regular $(n-1)$-simplex, connected by ${n \choose 2}$ identical springs.  We apply the representation theory of the symmetric group $S_n$ to compute its vibrational modes.
\end{abstract}

\thanks{C.H.C.\ was partially supported by Simons Foundation Collaboration Grant 519533.}
\thanks{This preprint has not undergone peer review (when
	applicable) or any post-submission improvements or corrections. The Version of Record of this
	article is published in \textit{The Mathematical Intelligencer}, and is available online at https://doi.org/10.1007/s00283-021-10160-z}

\maketitle

\thispagestyle{empty}

Take a collection of masses and connect them with springs.  You get a model of a classical molecule.  What are the vibrational modes of your ``molecule''?  This is an important physics problem.  One begins with the laws of Newton and Hooke and proceeds on a tour of some of the cornerstones of the undergraduate mathematics curriculum: differential equations, linear algebra, vector calculus, and, if the molecule has symmetries, abstract algebra.  Symmetries can be used to greatly reduce the computational difficulty of the problem---at the cost, of course, of a certain increase in conceptual difficulty!

Our purpose here is to take readers on this tour, and perhaps inspire some to guide their own students through it.  We must admit at the outset that the highlight of the tour, the use of symmetry, involves a subject beyond the scope of most undergraduate algebra courses: the representation theory of finite groups.  However, the necessary material can be built up from scratch for small groups, and the application to molecules provides a compelling introduction to the theory.

We will treat the maximally symmetric case of regular simplicial molecules with ``atoms'' of equal weight.  We begin with a review of the physical examples: ``dumbbells'', triangles, and tetrahedra.  Being mathematicians rather than physicists, we continue the sequence and find the vibrational frequencies of simplicial molecules of arbitrary dimension.  Although this result is perhaps impractical, it is, probably for that very reason, new, as far as we know.

At the end of the tour it will transpire that in fact, all the modes of the general solution are already exhibited by the tetrahedron.  This may come as a surprise to those of us uninitiated in the representations of the symmetric group, but as is so often the case, with hindsight one realizes one should have known it from the start.
%There are some features peculiar to the higher dimensional cases, but again, our main purpose is not so much to present original results as to advertise to mathematicians the beauty of the physical application.

\subsection*{History}
Here we consider classical molecules, but in quantum mechanics the same framework is relevant to infrared and Raman spectra.  Consequently, the topic is discussed by many authors.  Let us give a small selection.  On the mathematical side, Sternberg's 1994 text \cite{St94} gives a sophisticated treatment of the general theory, while on the physical side, Cracknell's 1968 text \cite{Cr68} takes a concrete approach.  Schrader's 1996 survey article \cite{Sc96} contains many references.  We also informally mention a few further physics and chemistry texts: Herzberg (1945), Wilson, Decius, and Cross (1955), Cotton (1963), Knox and Gold (1964), and Ezra (1983).  (And the triangular case makes an appearance as an exercise in Artin's algebra text.)

Looking through these, one is led to a 1936 survey article by Rosenthal and Murphy \cite{RM36}, a 1934 article by Wilson \cite{Wil34}, and to what appears to be the source of the ``best practice'' for handling symmetry, a 1930 article by Wigner \cite{Wig30} (a translation may be found in \cite{Cr68}).  Several authors, including Wigner himself and also Wilson, cite Brester's 1923 dissertation \cite{Br24} as a forerunner, but Wigner seems to have been the first to apply character theory in this context, a crucial advance.

We owe our own introduction to this problem to a text on mathematical methods in physics by Mathews and Walker~\cite{MW70}.  One of us once took a class taught from this book, which has a section on the triangular case.  The instructor covered it with great excitement, explaining how wonderful it is that you can use representation theory to avoid diagonalizing an enormous matrix.  He then assigned the tetrahedral case as homework.  Every single student attempted to solve it by diagonalizing an enormous matrix, and every single one got it wrong.  The author in question was never able to forget how terribly sad this made the instructor.  To make amends, three decades later he assigned the problem to the other author.

The students in the class were not alone in their fear of representation theory.  In the preface to \cite{St94}, Sternberg gives an account of the rise of symmetry in physics, including an amusing quote concerning the notorious ``gruppenpest''.  Wigner comments on the gradual acceptance of groups by physicists in the preface to the 1959 translation of his landmark 1931 book.  And Cracknell's delightful remark on Burnside's 1911 book in the guided bibliography of \cite{Cr68} shows that this acceptance could be strained by excessive abstraction.

\subsection*{The harmonic oscillator}
The tour begins with a spring.  Hooke's law states that there is a constant $k$ such that if the spring is stretched a distance~$x$ away from its equilibrium length, it pulls back with force $kx$.  Thus the ``spring force'' is $-kx$.  If $x$ is negative, the ``stretch'' is a compression, but the force is still $-kx$.

In our first differential equations course we encounter the following fundamental situation.  A mass~$m$ is placed in a horizontal frictionless trough.  One end of the spring is attached to the mass, the other to the end of the trough.  Only 1-dimensional motion is possible.  Newton tells us that the acceleration of the mass is proportional to the force on it.  We arrive at the harmonic oscillator equation
\begin{equation*}
   m \ddot x + kx = 0,
\end{equation*}
and we learn that its solutions have frequency $\omega_0 := \sqrt{k/m}$.

\subsection*{The dumbbell molecule}
Now put two equal masses in the trough, say $m_1$ and $m_2$, both of mass~$m$.  Connect them with the spring.  This models for example $\mathrm{O}_2$, oxygen in its breathable form.  Again, only 1-dimensional motion is possible.  Fix an equilibrium position of the system, and let $x_1$ and $x_2$ be the displacements of the masses from equilibrium.  The spring stretch is the difference $x_1 - x_2$, and so the forces on $m_1$ and $m_2$ are $\pm k(x_2 - x_1)$.  Thus we come to the vector differential equation
\begin{equation*}
   m \ddot x + kAx = 0,
   \qquad
   x := \begin{pmatrix} x_1 \\ x_2 \end{pmatrix},
   \qquad
   A := \begin{pmatrix} \phantom{-} 1 & -1 \\ -1 & \phantom{-} 1 \end{pmatrix}.
\end{equation*}

The ``pure mode'' solutions of this equation correspond to the eigenspaces of the matrix $A$, bringing us to linear algebra.  If $v$ is an eigenvector of eigenvalue $\lambda$, there is a pure mode along $v$ of frequency $\sqrt{\lambda}\, \omega_0$.

The eigenvalues of $A$ are $0$ and~$2$. The eigenvalue~$0$ corresponds to a translation of the entire molecule: the eigenvector is $(1, 1)$, and the two masses drift along together, experiencing no spring tension and never returning.

The eigenvalue $2$ has eigenvector $(1, -1)$, giving the only proper pure vibrational mode: the center of mass of the molecule remains at rest and the two masses move opposite to each other.

There is a puzzle for the student here.  With one mass, the frequency is $\omega_0$.  With two, it is $\sqrt{2}\, \omega_0$.  But should it not be that increasing the mass decreases the frequency?  The resolution is that in the one mass case, the trough plays the role of an infinite mass.  If the spring connects two unequal masses, say $m$ and $M$, the frequency turns out to be $\sqrt{k/\mu}$, where $\mu$ is the ``reduced mass'', $mM / (m+M)$.

\subsection*{The triangular molecule}
Take three masses, $m_1$, $m_2$, and $m_3$, all of mass~$m$.  Arrange them in an equilateral triangle and connect them by three identical springs, $k_{12}$, $k_{23}$, and $k_{13}$, all of spring constant~$k$.  Because sixty degree bond angles are extremely sharp, this configuration is rare in nature, but chemically inclined readers might look up the \textit{triatomic homonuclear molecules} cyclic ozone $\mathrm{O}_3$ (not yet synthesized), the helium trimer $\mathrm{He}_3$ (a not-at-all classical molecule forming scalene triangles), and the fascinating story of the trihydrogen cation $\mathrm{H}_3^+$ and its short-lived neutral cousin $\mathrm{H}_3$.

Restrict to 2-dimensional motion: place the molecule on a horizontal frictionless table.  Let $E_1$, $E_2$, and $E_3$ be the equilibrium positions of the masses, and let $x_1$, $x_2$, and $x_3$ be the displacements.  Thus for $i = 1$, $2$, or~$3$, $E_i$ is a constant 2-vector, $x_i$ is a time-dependent 2-vector, and the position of the mass $m_i$ at time~$t$ is $E_i + x_i(t)$.

The problem is no longer linear, and the exact equations of motion are quite formidable.  However, it is physically appropriate to make the ``small oscillation assumption'', taking the displacements to be small relative to the molecule.  Let $L$ be the rest length of the three springs: $L = |E_i - E_j|$ for all $i \not= j$.  Write $\approx$ for first order approximation in $x_i/L$.  A vector calculus exercise shows that the stretch in the spring $k_{ij}$ connecting $m_i$ to $m_j$ is a dot product:
\begin{equation*}
   \big| (E_i + x_i) - (E_j + x_j) \big| - \big| E_i - E_j \big| \approx
   L^{-1} (E_i - E_j) \cdot (x_i - x_j).
\end{equation*}

To obtain the vector force exerted by $k_{ij}$ on $m_i$, we must multiply this stretch by the unit vector from $m_i$ to $m_j$ and scale by~$k$.  The stretch is of first order, so we only need the unit vector to zeroth order: $(E_j - E_i) / L$.  The approximate force is thus
\begin{equation*}
   - k L^{-2} \big( (E_i - E_j) \cdot (x_i - x_j) \big) \big( E_i - E_j \big).
\end{equation*}
This is nothing other than $-k P_{ij} (x_i - x_j)$, where $P_{ij}$ is the $2 \times 2$ matrix projecting $\bR^2$ orthogonally to the line $\bR (E_i - E_j)$.  Note that $P_{ji} = P_{ij}$.

To obtain the total force on $m_i$, we must sum this over both values of $j$ not equal to~$i$.  This leads to a differential equation taking values in $\bR^6$.  It may again be written as $m \ddot x + kAx = 0$, where now $x$ and $A$ have block form:
\begin{equation*}
   x := \begin{pmatrix} x_1 \\ x_2 \\ x_3 \end{pmatrix},
   \qquad
   A := \begin{pmatrix} P_{12} + P_{13} & -P_{12} & -P_{13} \\ 
   -P_{12} & P_{12} + P_{23} & -P_{23} \\
   -P_{13} & -P_{23} & P_{13} + P_{23} \end{pmatrix}.
\end{equation*}

Just as for dumbbell molecules, the pure modes and their frequencies correspond to the eigenspaces and eigenvalues of $A$: eigenvectors of eigenvalue $\lambda$ give pure modes of frequency $\sqrt{\lambda}\, \omega_0$.  Because $P_{ij}$ is symmetric, $A$ is symmetric, and so it has an orthonormal basis of eigenvectors.  But it looks somewhat laborious to write $A$ down explicitly and diagonalize it, so for now we will only make a few easily checked observations:

\begin{itemize}

\item
If all three $x_i$ are equal, then $x$ is an eigenvector of eigenvalue~$0$.  These trivial solutions are translations of the entire molecule, analogous to the eigenvalue~$0$ solution of the dumbbell.  

\smallbreak \item
Let $R$ be a $2 \times 2$ matrix rotating $\bR^2$ by $90$~degrees.  Using $P_{ij} R (E_i - E_j) = 0$, we find that $(R E_1, R E_2, R E_3)$ is another eigenvector of eigenvalue~$0$.  It corresponds to rotation of the entire molecule.

Like translation, rotation is not a proper vibration, as the $x_i$ do not remain small.  The matrix $R$ enters the picture because it is an ``infinitesimal rotation'': to first order, rotation by a small angle $\varepsilon$ is $I + \varepsilon R$.

\smallbreak \item
Place the center of mass of the molecule at the origin, so that $E_1 + E_2 + E_3 = 0$.  Then using $P_{ij} (E_i - E_j) = E_i - E_j$, we find that $(E_1, E_2, E_3)$ is an eigenvector of eigenvalue~$3$.  It corresponds to the most intuitively obvious proper vibrational mode, the so-called ``breathing'' mode, in which all three masses move towards and away from the origin synchronously.

\end{itemize}

These modes account for four orthonormal eigenvectors of $A$, of eigenvalues $0$, $0$, $0$, and~$3$.  What are the other two?  Visualize the triangle in the ``upright'' position.  One might guess that there is a solution in which the mass at the top goes upward while the others go downward and inward.  But would there not be three such solutions, one for each vertex?  And can we deduce their frequencies without work?

This guess will turn out to be correct, and there are indeed three such solutions, but the three corresponding eigenvectors span only a 2-dimensional space: they sum to zero.  And yes, we can deduce their frequencies: clearly the two new eigenvalues are the same, say $\alpha$, so the trace of $A$ is $3 + 2\alpha$, the sum of its eigenvalues.  But the trace of any rank~1 projection is~$1$, so the trace of $A$ is~$6$.  Thus $\alpha = 3/2$.

At this point one begins to see the role of abstract algebra, specifically, groups.  It is physically clear that the symmetry group of the triangle, the dihedral group $D_3$, must transform any solution of our differential equation into another solution.  Put differently, $D_3$ acts on the 6-dimensional vector space of vibrational motions of the triangular molecule.  Because the equation is linear, the action is linear.  Such actions are called \textit{representations} of $D_3$.

It is also physically clear that $D_3$'s action must take any pure mode solution to another of the same frequency.  Reviewing the discussion, we see that the action partitions the six eigenvalues of $A$ into four subsets: $\{0, 0\}$, $\{0\}$, $\{3\}$, and $\{3/2, 3/2\}$.  These subsets will turn out to correspond to \textit{irreducible subrepresentations.}

\subsection*{The tetrahedral molecule}
Arrange four identical masses $m_i$ in a regular tetrahedron and connect them by six identical springs $k_{ij}$, where $1 \le i,\, j \le 4$.  In the real world such molecules are rare, and not to be confused with tetrahedral molecules of the more common $\mathrm{AB}_4$ type such as methane, which have one central atom bonded to four identical surrounding atoms.  However, they do occur: yellow arsenic, $\mathrm{As}_4$, is the most toxic form of arsenic, and white phosphorus, $\mathrm{P}_4$, is another homonuclear molecule with an unpleasant reputation.

We must now allow 3-dimensional motion, so instead of placing our model on a frictionless table, we must bring it with us on a space ship.  (We could obtain an approximation by hanging it from the ceiling, but then gravity would differentiate the spring lengths.)  For $1 \le i \le 4$, let $E_i$ and $x_i(t)$ be the equilibrium positions and displacements of the masses.  These are now 3-vectors, but nevertheless everything goes exactly as for triangular molecules: we still have $|E_i - E_j| = L$ for $i \not= j$, and the same argument shows that to first order, the force exerted by the spring $k_{ij}$ on the mass $m_i$ is $-k P_{ij} (x_i - x_j)$, where $P_{ij}$ is the $3 \times 3$ matrix projecting $\bR^3$ orthogonally to the line $\bR (E_i - E_j)$.

The total force on $m_i$ is the sum of the forces exerted by the three springs attached to it, giving
\begin{equation} \label{Fi}
   m \ddot x_i = -k \sum_{j \not= i} P_{ij} (x_i - x_j).
\end{equation}
Proceeding as before, let $x(t)$ be the 12-vector obtained by stacking up the $x_i$.  Once again we obtain the equation $m \ddot x + kAx = 0$, where now $A$ is a $12 \times 12$ matrix with $3 \times 3$ block entries $A_{ij}$ for $1 \le i, j \le 4$, given by
\begin{equation} \label{Aij}
   A_{ij} = -P_{ij}\ \text{ for }\ i \not= j,
   \qquad
   A_{ii} = \sum_{j \not= i} P_{ij}.
\end{equation}

Again, the pure modes and their frequencies correspond to the diagonalization of $A$.  But here it is downright daunting to write $A$ explicitly, let alone diagonalize it.  The observations from the triangular case do carry over; they no longer permit us to guess all the frequencies, but even so it is useful to restate them:

\begin{itemize}

\item
If all four $x_i$ are equal, then $x$ is an eigenvector of eigenvalue~$0$, giving three trivial translation solutions.

\smallbreak \item
Let $R$ be any real skew-symmetric $3 \times 3$ matrix.  Then $y \cdot Ry = 0$ for all $y \in \bR^3$, and so again we have $P_{ij} R (E_i - E_j) = 0$, implying that $(R E_1, R E_2, R E_3, R E_4)$ is also an eigenvector of eigenvalue~$0$.

As before, these eigenvectors correspond to rotations of the entire molecule, and $R$ is an infinitesimal rotation.  In fact, the set of all such $R$ is the Lie algebra of the Lie group $\mathrm{SO}_3$ of rotations of space.

\smallbreak \item
Placing the center of mass of the molecule at the origin so as to have $\sum_1^4 E_i = 0$, we again find that $(E_1, E_2, E_3, E_4)$ is an eigenvector.  It still corresponds to the breathing mode, but here it has eigenvalue~$4$.

\end{itemize}

The reader may verify that these modes account for seven of the twelve eigenvalues of $A$.  What are the other five?  It is entertaining to try to guess the general forms of their modes.  Our earlier reasoning concerning the trace of $A$ shows that the sum of all twelve eigenvalues is~$12$.  Therefore the sum of the five unknown eigenvalues must be~$8$, but we cannot yet say more.

What does a preliminary consideration of symmetry tell us?  First, it is important to understand that the full symmetry group is the symmetric group $S_4$, which acts on the tetrahedron by reflections and reflection-rotations as well as rotations.  One might think that only the alternating group $A_4$, which acts solely by rotations, should be allowed, because reflection symmetries cannot be enacted by moving the molecule in 3-space.  But this would be incorrect: both reflection and rotation symmetries preserve solutions of the equation.

Thus the action of the symmetry group on the solution space is a 12-dimensional representation of $S_4$.  As before, it preserves the frequencies of pure modes, which leads to a partitioning of the eigenvalues of $A$.  The seven eigenvalues we have encountered so far make up three of the partition subsets: $\{0, 0, 0\}$, $\{0, 0, 0\}$, and $\{4\}$.  In order to tell the rest of the story, let us turn to the general case and present Wigner's method.

\subsection*{The simplicial molecule}
Fix a positive integer~$n$.  Arrange $n$ masses, all of mass~$m$, in a regular simplex.  Connect them by $\binom{n}{2}$ springs, all of spring constant~$k$ and length~$L$.  This molecule occupies an $(n-1)$-dimensional space, so for $n \ge 5$ it is not physical, but the laws of Hooke and Newton still make sense.  Label the masses $m_1, \ldots, m_n$, and let $E_i$ and $x_i(t)$ be the equilibrium position and displacement of $m_i$.  Demonstrate as before that to first order, the spring connecting $m_i$ and $m_j$ exerts a force $-k P_{ij} (x_i - x_j)$ on $m_i$, where $P_{ij}$ is orthogonal projection to $\bR (E_i - E_j)$.

Summing these forces over $j \not= i$ shows that the total force on $m_i$ is still given by~\eqref{Fi}.  This gives the usual equation,
\begin{equation} \label{HN}
   m \ddot x + k A x = 0,
\end{equation}
where $x$ is the block $n$-vector whose $i^\thup$ entry is $x_i$, and $A$ is the block $n \times n$ matrix with entries~\eqref{Aij}.  

Let us step back and take stock of the situation.  The pure mode solutions of~\eqref{HN} are $x(t) = v \exp(\pm i \omega_0 t \sqrt{\lambda})$, where $v$ is an eigenvector of $A$ of eigenvalue $\lambda$.  The symmetry group of the molecule is the symmetric group $S_n$ of all permutations of the equilibrium positions~$E_i$.  It acts orthogonally on the ambient space of the molecule, and this action defines a linear action on the space of all solutions of~\eqref{HN} which maps pure mode solutions to others of the same frequency.  

It will be helpful to rephrase these observations.  Write $V$ for the space of column vectors~$x$, the set of all possible initial conditions for~\eqref{HN}.  The action of $S_n$ on the solution space is time-independent, and so it is really a representation on $V$, i.e., a group homomorphism $\phi: S_n \to \mathrm{GL}(V)$.  If $x(t)$ is a solution, then for all $\sigma$ in $S_n$, $\phi(\sigma) x(t)$ is also a solution.

Let $V_\lambda$ be the $\lambda$-eigenspace of $A$, the set of initial conditions giving the pure mode of frequency $\omega_0 \sqrt{\lambda}$.  Because $A$ is symmetric, $V = \bigoplus_\lambda^\perp V_\lambda$.  The statement that the action preserves frequencies may be recast in the following three equivalent ways:

\begin{itemize}

\item
For all $\sigma \in S_n$ and $\lambda \in \bC$, $\phi(\sigma)$ preserves $V_\lambda$.

\smallbreak \item
For all $\sigma \in S_n$, $\phi(\sigma)$ and $A$ commute.

\smallbreak \item
For all $\lambda \in \bC$, $V_\lambda$ is a subrepresentation of $V$.

\end{itemize}

\subsection*{Wigner's method}
The idea is to use the \textit{character} of $\phi$.  The character of a representation is simply its trace.  For example, the character of $\phi$ is the function
\begin{equation*}
   \tr(\phi): S_n \to \bR,
   \qquad
   \sigma \mapsto \tr\big(\phi(\sigma)\big).
\end{equation*}
The character theory of finite and even compact groups had been well developed by 1930, and it was known that the representations of such groups are completely determined by their characters, up to the natural notion of equivalence.  This is at first sight amazing, because of course the trace of a matrix does not determine its similarity class.  To get a glimpse of why it might be true, remember that a representation is a homomorphism.  Therefore, if a matrix is in its image, so are all the powers of that matrix.  And the traces of all the powers of a diagonalizable matrix do determine its similarity class!

Not only do the characters determine the representations, they do so in a computationally effective way, via the \textit{orthogonality relations} for the irreducible representations.  These relations are the gateway to many modern introductions to representation theory, such as \cite{FH91}.

Because $V$ is the direct sum of its subrepresentations $V_\lambda$, its character is the sum of their characters:
\begin{equation*}
   \tr\big(\phi(\sigma)\big) = \sum_\lambda \tr\big(\phi(\sigma)|_{V_\lambda}\big).
\end{equation*}
Because $V_\lambda$ is the $\lambda$-eigenspace of $A$, we have
\begin{equation} \label{W}
   \tr\big(A \phi(\sigma)\big) = \sum_\lambda \lambda\, \tr\big(\phi(\sigma)|_{V_\lambda}\big).
\end{equation}
This is effectively a system of linear equations in the eigenvalues~$\lambda$, one equation for each $\sigma$ in $S_n$.  Wigner says ``just solve the system!''

It remains only to ``unpack'' these somewhat cryptic instructions.  Up to this point we have assiduously avoided making explicit choices of the equilibrium positions $E_i$, but it is now time to face this task.  Following our discussion of the $n = 2$, $3$, and~$4$ cases, one expects $E_i$ and $x_i$ to be in $\bR^{n-1}$, and indeed the reader may have wondered why we wrote $V$ for the space of vectors~$x$, rather than simply $\bR^{n(n-1)}$.  The reason is that we will situate our molecule not in $\bR^{n-1}$, but $\bR^n$!  Why complicate matters by adding a dimension?  There is a standard trick: the simplest choice of the vertices of a regular $(n-1)$-simplex is the standard basis $e_1, \ldots, e_n$ of $\bR^n$.  Thus we set $E_i := e_i$ and permit $x_i$ to vary in $\bR^n$.

\subsection*{Tensors}
The next step is to bring in the language of tensor products.  For students unfamiliar with them, the \textit{Kronecker product} makes them concrete: given matrices $M$ and $M'$ of any sizes whatsoever, $M \ot M'$ is the block matrix with $ij^\thup$ entry $M_{ij} M'$.  We regard $V$ as $\bR^n \ot \bR^n$ and make the identifications
\begin{equation*}
   x = \sum_{i=1}^n e_i \ot x_i,
   \qquad
   A = \sum_{i, j = 1}^n (e_i e_j^T) \otimes A_{ij},
\end{equation*}
where $T$ denotes transpose and hence $e_i e_j^T$ is the $n \times n$ matrix with a $1$ in the $ij^\thup$ entry and $0$'s elsewhere.

This puts us in position to give formulas for $A$ and $\phi$.  Here is an exercise: the projection $P_{ij}$ is $\frac{1}{2} (e_i - e_j) (e_i - e_j)^T$, and \eqref{Aij} may be rewritten (quite elegantly!) as
\begin{equation} \label{A}
   A = \sum_{i, j = 1}^n (e_i e_i^T - e_i e_j^T) \otimes P_{ij}
   = \sum_{i, j = 1}^n P_{ij} \otimes P_{ij}.
\end{equation}

In order to understand $\phi$, consider the action of $S_n$ on the molecule and its ambient space $\bR^n$.  It permutes the equilibrium positions of the atoms, and so it is the usual permutation representation $\pi: S_n \to \mathrm{GL}_n \bR$ permuting the standard basis vectors: $\pi(\sigma) e_i := e_{\sigma(i)}$.

How does $\phi$ act on a displacement vector $x$ in $\bR^n \ot \bR^n$?  The displacement $x_i$ of the mass $m_i$ in the configuration $x$ determines the displacement $(\phi(\sigma) x)_{\sigma(i)}$ of the mass $m_{\sigma(i)}$ in the configuration $\phi(\sigma) x$.  However, the two are not equal, because the same orthogonal motion of $\bR^n$ that permutes the equilibrium positions also acts on the displacements.  Therefore $(\phi(\sigma) x)_{\sigma(i)}$ is equal to $\pi(\sigma) x_i$.  All this can be stated succinctly in a single equation: for $\sigma \in S_n$,
\begin{equation*}
   \phi(\sigma) = \pi(\sigma) \ot \pi(\sigma).
\end{equation*}
One says that $\phi$ is $\pi \ot \pi$, or $\ott \pi$, the \textit{tensor square} of $\pi$.  The symmetry of~\eqref{A} makes it simple to verify the commutativity of $\ot^2 \pi(\sigma)$ and $A$ directly.

At this stage we require two facts, which enterprising students will easily prove:

\begin{itemize}

\item
$\bR^n \ot \bR^n$ is $\cS^2 \bR^n \oplus \Lambda^2 \bR^n$, the sum of the \textit{symmetric} and \textit{alternating} squares:
\begin{align*}
   & \cS^2 \bR^n := \Span \big\{ y \ot z + z \ot y: y, z \in \bR^n \big\}, \\
   & \Lambda^2 \bR^n := \Span \big\{ y \ot z - z \ot y: y, z \in \bR^n \big\}.
\end{align*}
These spaces are mutually orthogonal subrepresentations under $\ott \pi$.

\smallbreak \item
$\bR^n = \bR f \oplus f^\perp$, where $f := \frac{1}{n} \sum_i e_i$ and $f^\perp$ is its orthogonal complement.  Both summands are subrepresentations under $\pi$: the line $\bR f$ is a trivial representation, orthogonal to all the edges $e_i - e_j$ of the molecule, and $f^\perp$ is the hyperplane spanned by the vectors $e_i - f$, a translation of the space in which the molecule sits.

\end{itemize}

Following our pattern, we begin the process of finding the eigenspaces of $A$ by treating translations and rotations:

\begin{itemize}

\item
The space of translations is $f \ot \bR^n$, the vectors $f \ot ny$ with all $x_i$ equal to~$y$.  Using~\eqref{A} and $P_{ij} f = 0$, one checks that they have $A$-eigenvalue $0$.

\smallbreak \item
The space of infinitesimal rotations is $\Lambda^2 \bR^n$.  By~\eqref{A}, its elements have $A$-eigenvalue $0$.  They may be written in the form $\sum_i e_i \ot R e_i$, where $R$ is a skew-symmetric matrix, i.e., an element of the Lie algebra of $\mathrm{SO}_n$.

\end{itemize}

Some remarks are in order.  The decomposition $\bR^n = \bR f \oplus f^\perp$ yields orthogonal decompositions of the various tensor spaces involving $\bR^n$:
\begin{subequations} \label{plethyism}
\begin{align}
   \label{ott} & \ott(\bR f \oplus f^\perp) =
   \bR (f \ot f) \oplus (f \ot f^\perp) \oplus (f^\perp \ot f) \oplus (\ott f^\perp), \\
   \label{sym} & \cS^2 (\bR f \oplus f^\perp) =
   \bR (f \ot f) \oplus \Span \big\{f \ot y + y \ot f: y \in f^\perp \big\} \oplus \cS^2 f^\perp, \\
   \label{alt} & \Lambda^2 (\bR f \oplus f^\perp) =
   \Span \big\{f \ot y - y \ot f: y \in f^\perp \big\} \oplus \Lambda^2 f^\perp.
\end{align}
\end{subequations}

The translation space $f \ot \bR^n$ is the sum $\bR (f \ot f) \oplus (f \ot f^\perp)$ of the first two summands of~\eqref{ott}.  The former is the translations along $f$, a trivial subrepresentation, and the latter is the translations in the space of the molecule, a copy of $f^\perp$.

The rotation space is~\eqref{alt}.  The first summand gives rotations in planes containing~$f$; it is a copy of $f^\perp$.  The second gives rotations in planes in $f^\perp$.

Combine these to see that the sum of the translation and rotation spaces is
\begin{equation*}
      \bR (f \ot f) \oplus (f \ot f^\perp) \oplus (f^\perp \ot f) \oplus (\Lambda^2 f^\perp).
\end{equation*}
This is the orthogonal complement of $\cS^2 f^\perp$.  Thus, without any computation, we arrive at the following statement:
\begin{itemize}

\item[] \textit{
The $A$-eigenspaces $V_\lambda$ with $\lambda \not= 0$ are all subrepresentations of $\cS^2 f^\perp$.}

\end{itemize}

Before proceeding we require one more fact, which even enterprising students may have trouble proving:

\begin{itemize}

\item[] \textit{
$\cS^2 f^\perp$ contains exactly three irreducible subrepresentations: a trivial 1-dimensional representation $B$, a copy of $f^\perp$ which we will call $C$, and a representation $D$ of a new type.  It is equal to their direct sum:}
\begin{equation*}
   \cS^2 f^\perp = B \oplus C \oplus D.
\end{equation*}

\end{itemize}
Let us indicate how a reader of \cite{FH91} might reach this result in minimal time: having mastered Lectures~1 and~2, apply the same strategy used in \S3.2 to prove that $f^\perp$ and $\Lambda^2 f^\perp$ are irreducible (compare to Exercise~4.19).

The spaces $B$, $C$, and $D$ are in fact the pure modes of the molecule!  Why is this?  Each $A$-eigenspace $V_\lambda$ of non-zero eigenvalue $\lambda$, being contained in $\cS^2 f^\perp$, must be a direct sum of irreducible subrepresentations of $\cS^2 f^\perp$.  And $B$, $C$, and $D$ are the only possibilities!  Turn this statement around to say that $A$ acts by scalars on $B$, $C$, and $D$, and write $\lambda_B$, $\lambda_C$, and $\lambda_D$ for the scalars.

\subsection*{Characters}
At last we are prepared to take up Wigner's equation~\eqref{W}.  Consider first its right hand side.  The sum is really only over non-zero eigenvalues~$\lambda$, and so it reduces to
\begin{equation} \label{BCD}
   \lambda_B\, \tr \big(\ott \pi|_B \big) +
   \lambda_C\, \tr \big(\ott \pi|_C \big) +
   \lambda_D\, \tr \big(\ott \pi|_D \big).
\end{equation}

It is time to compute these characters.  Given a permutation $\sigma$, let $\iota_k(\sigma)$ denote the number of $k$-cycles in its disjoint cycle decomposition.  Some easy exercises:

\begin{itemize}

\smallbreak \item
$\tr \big(\ott \pi|_B \big)$ is simply the constant function~$1$ on $S_n$, as $B$ is trivial.

\medbreak \item
$\tr \big(\ott \pi|_C \big) = \tr \big(\pi|_{f^\perp} \big)$, as $C$ is a copy, i.e., a conjugate, of $f^\perp$.

\medbreak \item
$\tr \big(\pi|_{f^\perp} \big) + 1 = \tr \big( \pi \big)$, as $f^\perp \oplus \bR f = \bR^n$.

\medbreak \item
$\tr \big( \pi \big) = \iota_1$, i.e., $\tr \big( \pi(\sigma) \big)$ is the number of points $\sigma$ fixes.

\end{itemize}

Thus the characters of $B$ and $C$ are $1$ and $\iota_1 - 1$.  How will we find the character of $D$?  Clearly the characters of $B$, $C$, and $D$ sum to the character of $\cS^2 \bR^n$, but the latter can be difficult to understand.  Here is an opaquely phrased hint: if $\{ y_i \}$ is an $M$-eigenbasis, then both $\{ y_i \ot y_j \}$ and $\{ y_i \ot y_j \pm y_j \ot y_i \}$ are $(\ott M)$-eigenbases.  Use it and \eqref{sym} to do a few more exercises (compare to Exercise~4.15 of~\cite{FH91}):

\begin{itemize}

\smallbreak \item
$\tr(\ott \pi) = \tr^2(\pi) = \iota_1^2$.

\medbreak \item
$\tr \big( \ott \pi|_{\cS^2 \bR^n} \big) =
\frac{1}{2} \big[ \tr^2 \big( \pi(\sigma) \big) + \tr \big( \pi(\sigma^2) \big) \big] =
\frac{1}{2} \big[ \iota_1^2 + (\iota_1 + 2 \iota_2) \big]$.

\medbreak \item
$\tr \big(\ott \pi|_D \big) = \frac{1}{2} \iota_1 (\iota_1  - 3) + \iota_2$.

\end{itemize}

We have tamed the right hand side of~\eqref{W}: \eqref{BCD} has become
\begin{equation} \label{RHS}
   \lambda_B +
   \lambda_C \big( \iota_1 - 1 \big) +
   \lambda_D \big( {\textstyle\frac{1}{2}} \iota_1 (\iota_1 - 3) + \iota_2 \big).
\end{equation}

Now consider the left hand side.  This time let us begin by stating the end result:
\begin{equation} \label{LHS}
   \tr \big( A \circ \ott \pi \big) = n +
   {\textstyle \frac{1}{2}} n \big( \iota_1 - 1 \big) +
   \big( {\textstyle \frac{1}{2}} \iota_1 (\iota_1  - 3) + \iota_2 \big).
\end{equation}
Here is a series of hints leading to the proof:
\begin{itemize}

\medbreak \item
$\tr \sum_{ij} \ott \big(P_{ij}\, \pi(\sigma) \big) =
\sum_{ij} \tr^2 \big(P_{ij}\, \pi(\sigma) \big)$.

\medbreak \item
$P_{ij}\, \pi(\sigma) e_k =
\frac{1}{2} (\delta_{i, \sigma(k)} - \delta_{j, \sigma(k)}) (e_i - e_j)$, and so
\begin{equation*}
   \tr \big(P_{ij}\, \pi(\sigma) \big) =
   {\textstyle \frac{1}{2}} \big( \delta_{i, \sigma(i)} - \delta_{i, \sigma(j)} -
   \delta_{j, \sigma(i)} + \delta_{j, \sigma(j)} \big).
\end{equation*}

\medbreak \item
The sum is really over $i \not= j$, so in the square of the trace, the cross-terms
$\delta_{i, \sigma(i)} \delta_{i, \sigma(j)}$, $\delta_{j, \sigma(i)} \delta_{j, \sigma(j)}$,
$\delta_{i, \sigma(i)} \delta_{j, \sigma(i)}$, and $\delta_{i, \sigma(j)} \delta_{j, \sigma(j)}$
contribute~$0$.

\medbreak \item
$ \tr \big( A \circ \ott \pi(\sigma) \big) = \frac{1}{2} \sum_{i \not= j}
\big( \delta_{i, \sigma(i)} + \delta_{i, \sigma(j)} +
\delta_{i, \sigma(i)} \delta_{j, \sigma(j)} +
\delta_{i, \sigma(j)} \delta_{j, \sigma(i)} \big). $

\medbreak \item
These summands contribute $(n-1) \iota_1$, $(n - \iota_1)$, $\iota_1 (\iota_1 - 1)$, and $2 \iota_2$.

\end{itemize}

We are done!  Comparing \eqref{RHS} with \eqref{LHS}, we have the eigenvalues of the modes:
\begin{equation*}
   \lambda_B = n, \qquad
   \lambda_C = {\textstyle\frac{1}{2}} n, \qquad
   \lambda_D = 1.
\end{equation*}

\subsection*{Die Gruppenpest}
What do these modes look like?  And could we have found them without ``that pesty group business'', as Wigner translated the phrase?  (See his \textit{Recollections}, as told very touchingly by Szanton.)

The trivial representation $B$ is the breathing mode.  It is spanned by the vector $f_B := \frac{1}{n} \sum_i e_i \ot (e_i - f)$, because $e_i - f$ points radially outward from the center of the molecule to the mass $m_i$.  We know that $B$ is in $\cS^2 f^\perp$, and indeed $f_B$ is equal to $\frac{1}{n} \sum_i \ott(e_i - f)$.  Direct computation using \eqref{A} gives $\lambda_B = n$.

The copy $C$ of $f^\perp$ is also easy to find.  The vectors $\ott(e_i - f)$ clearly span an $n$-dimensional subrepresentation of $\cS^2 f^\perp$ which is isomorphic to the permutation representation $\bR^n$ by $e_i \mapsto \ott(e_i - f)$.  Hence $C$ is the $(n-1)$-dimensional subspace spanned by the vectors $\ott(e_i - f) - f_B$.  Again, \eqref{A} gives $\lambda_C = \frac{1}{2} n$ directly.

The vibrations corresponding to these spanning vectors of $C$ are just what one would guess: as the mass $m_i$ moves directly away from the center, all the other masses move away from the center in the opposite direction and inward toward the line through $m_i$ and the center.  

And now $\lambda_D$ follows simply from the trace of $A$ itself!  So is it possible after all to avoid groups?  Of course, even if it were, it would take nothing away from Wigner's method, which applies to \textit{every} molecule with symmetry.  But wait: without Wigner, we cannot say ``$\lambda_D$ follows'', because as we said in the introduction, we would not know that there are only three modes.  Without knowing the decomposition of $\cS^2 f^\perp$ under $S_n$, one might be excused for expecting more and more modes as $n$ grows.

A persistent avoider of groups might try to find a way around this.  Having understood the translations and rotations and the modes $B$ and $C$, one knows that the orthogonal complement in $\ott \bR^n$ of their sum must make up the remaining modes.  This complement turns out to be the space
\begin{equation} \label{D}
  \left\{ {\textstyle \sum_{ij}}\, x_{ij}\, (e_i - f) \ot (e_j - f):
  x_{ij} = x_{ji},\ x_{ii} = 0,\ {\textstyle \sum_k x_{ik}} = 0\ \forall\ i, j \right\}
\end{equation}
(let us note that under these conditions, the sum reduces to $\sum_{ij} x_{ij}\, e_i \ot e_j$).

Applying \eqref{A}, we find that $A$ acts on \eqref{D} by~$1$, i.e., \eqref{D} is the 1-eigenspace of $A$.  Does this solve the problem without groups?  Not yet: it could still be that there is more than one mode of eigenvalue~$1$, or in group-theoretic terms, that \eqref{D} decomposes as a sum of more than one irreducible representation of $S_n$.

The description \eqref{D} of $D$ yields a curious fact: given a $D$-vibration of the $(n-1)$-simplex, we can obtain a $D$-vibration of the $n$-simplex by adding a mass at rest.  Because $D$ first appears at $n = 4$, the 3-dimensional case, this phenomenon does not arise for physical molecules, but for $n \ge 5$ there are $D$-vibrations in which only four masses move, built by adding masses at rest to tetrahedral $D$-vibrations.  These vibrations span $D$ and $S_n$ acts transitively on them, but even this does not quite show that $D$ is irreducible.

The typical tetrahedral $D$-vibration consists of two pairs of masses moving in opposition, the masses in each pair going towards each other as the pair moves away from the center.  The $(n-1)$-simplex has $\frac{1}{2} \binom{n}{2} \binom{n-2}{2}$ such vibrations, which can averaged in various ways to produce other $D$-vibrations.  For example, averaging over all $(2, n-2)$ partitions gives an $S_n$-invariant spanning set with $\binom{n}{2}$ elements, only $n$ more than the dimension of $D$.  Puzzle: is this the smallest $S_n$-invariant spanning set?

Let us conclude with a remark for those who would rather embrace groups than avoid them: $S_n$ is not the full symmetry group of the state space!  The exchange involution $s$ sending $y \ot z$ to $z \ot y$ clearly commutes with both \eqref{A} and $\ott \pi$, extending the action of $S_n$ on $\ott \bR^n$ to an action of $S_n \times \bZ_2$.  Another puzzle: if $x(t)$ is a solution of \eqref{HN}, explain physically why is $s \big( x(t) \big)$ also a solution.

% Let us conclude with a remark for those who would rather embrace groups than avoid them: S_n is not the full group of solution-preserving transformations! The exchange involution... with both (5) and \otimes^2 \pi, making the solution space a representation of S_n \times Z^2. We also ...

\end{document}